\documentclass[epj]{svjour}
\usepackage{graphics}
\usepackage{dcolumn}
\usepackage{amsmath}
\usepackage{amssymb,amsfonts,latexsym}
\usepackage{bm}
\usepackage[mathcal]{euscript}
\usepackage{graphicx}
\usepackage{epsfig}
\usepackage{color}

\newcommand{\be}{\begin{equation}}
\newcommand{\ee}{\end{equation}}
\newcommand{\ben}{\begin{equation*}}
\newcommand{\een}{\end{equation*}}
\newcommand{\bea}{\begin{eqnarray}}
\newcommand{\eea}{\end{eqnarray}}

\newcommand \ve {\varepsilon}
\begin{document}
\graphicspath{{../figures/}}
\title{The glass transition in a nutshell: a source of inspiration to describe the subcritical transition to turbulence}
\titlerunning{Turbulence and Glass transitions}
\author{Olivier Dauchot\inst{1} \and Eric Bertin\inst{2}}
\institute{EC2M, ESPCI-ParisTech, UMR Gulliver 7083 CNRS, 75005 Paris, France \and Universit\'e de Lyon, Laboratoire de Physique, ENS Lyon, CNRS, 46 All\'ee d'Italie, 69007 Lyon, France}
\date{Received: date / Revised version: date}
\abstract{
The starting point of the present work is the observation of possible analogies, both at the phenomenological and at the methodological level, between the subcritical transition to turbulence and the glass transition. Having recalled the phenomenology of the subcritical transition to turbulence, we review the theories of the glass transition at a very basic level, focusing on the history of their development as well as on the concepts they have elaborated. Doing so, we aim at attracting the attention on the above mentioned analogies, which we believe could inspire new developments in the theory of the subcritical transition to turbulence. We then briefly describe a model inspired by one of the simplest and most inspiring model of the glass transition, the so-called Random Energy Model, as a first step in that direction.
} 
\maketitle
\section{Introduction}
\label{intro}
In a recent work~\cite{Dauchot2012}, we discussed possible analogies between the subcritical transition to turbulence in shear flows and the glass transition in supercooled liquids. While no precise mapping between the glass transition and the transition to turbulence should be expected,  there are similarities at the phenomenological and methodological levels, which suggest that inspiring cross-fertilization could arise from a closer inspection of both phenomenologies as well as from the history of the glass transition theories, which have developed a lot in the last half-century. 

The aim of~\cite{Dauchot2012} was to call the attention of the statistical physics community on  the old standing problem of the transition to turbulence. Accordingly, we presented a rather complete review of the experimental results of the subcritical transition, focusing on the recent surge of interest regarding the statistics of the turbulent lifetimes, while assuming a minimal knowledge of the glass transition. Here, we shall follow the opposite viewpoint. We aim at providing a basic introduction to the glass transition and its theories, focusing on the history of its development as well as on the concepts it has elaborated. Our motivation is to answer those hydrodynamicist colleagues and friends, who, after our oral presentation at the colloquium in honor of Paul Manneville [present special issue of EPJE], had some doubts about the possibility to enter the meanders of the glass transition literature without some guidebook. Accordingly, we have deliberately adopted a narrative style, emphasizing ideas and concepts, and referring to the original publications for the more technically involved discussions.

The paper is organized as follows. After a brief review of the phenomenology of the subcritical transition to turbulence and a few remarks about its theoretical understanding, we come to our pedagogical introduction to the theory of the glass transition. One serious difficulty is that there is no general agreement on \emph{what is} the theory of the glass transition. Another one is that none of the present authors has a complete view of the field, with all its subtleties and related debates which make the scientific meetings in that community so vivid. We thus have chosen to tell a simple but inevitably partial story, which reflects our own understanding. The good counterpart is that there is a chance that it will be accessible to those who are not already familiar with it. And we hope that it will be complete enough to draw some lines along which, we believe, some conceptual progress could be done regarding the subcritical transition to turbulence. We finally recast the elements and the results of a very simple model inspired by the so-called Random Energy Model~\cite{Derrida1980}, which we have proposed as a first step along this (long) path~\cite{Dauchot2012}.

\section{Subcritical transition to turbulence}
\label{sec:1}
The subcritical transition to turbulence occurs in flow regimes lacking linear instability and referred to as globally subcritical~\cite{Joseph1976,Grossmann2000,Dauchot1997}. Plane Couette flow (pCf), driven by two plane walls moving parallel to each other in opposite directions, is linearly stable at all Reynolds number and as such is the epitome of globally subcritical transitions~\cite{Romanov1973,Drazin1981}. Circular Poiseuille flow (cPf), the flow driven by a pressure gradient in an axisymmetric pipe, is believed to share the same property, though this, to date, has not been formally proven. 

\subsection{A brief review of the phenomenology}
\label{subsec:1-1}

The essential features of the subcritical transition to turbulence are as follow.
\begin{itemize}
\item \emph{Subcriticality}: Finite amplitude perturbations of the flow may yield an abrupt transition, although the laminar state is stable against infinitesimal perturbations. A critical amplitude of perturbation can thus be defined, meaning that the flow is globally subcritical.
Such finite amplitude perturbations may however lead to very different states of the flow. At the same Reynolds number, one observes either a relaxation to the laminar state, or a sustained disordered flow, composed of coexisting laminar and turbulent domains. The latter case is observed in particular by suddenly lowering the Reynolds number from a value where the flow is fully turbulent to a value in the transitional regime (a procedure called a ``quench'' in the following).

\item \emph{Spatiotemporal intermittency}: 
The turbulent patches forming the disordered flow are observed to move, grow and decay, split and merge, leading to a coexistence dynamics called spatiotemporal intermittency, in which turbulent (``active'') regions invade laminar (``absorbing'') ones where turbulence cannot emerge in a spontaneous way.

\item \emph{Transients and metastability}:
There are two ways to prepare the system in a non laminar state. Either one sets the Reynolds number in the transitional regime and creates a turbulent patch from a finite amplitude localized perturbation, or one quenches the flow from a homogeneous turbulent state prepared at high Reynolds number. In the first case, one measures the lifetime of a single turbulent patch. In the second case, one measures the lifetime of the spatially extended disordered state. In the cPf, only the first procedure has been followed; both were used in the pCf case.  The distributions of these lifetimes (obtained by repeating many times the same experiment) are exponential and in both cases, the corresponding characteristic time $\tau$ increases with Reynolds number.
An open issue, which is still a matter of debate, is to know whether $\tau$ diverges or not at a finite value $R_g$ of the Reynolds number. Note that below a value $R_u$, the flow relaxes rapidly to the laminar state.

\item \emph{Unstable states}:
Numerous unstable solutions appear in phase space when the lifetime of the disordered flow becomes large. All, or nearly all, of these unstable states that are known for these flows are periodic orbits. Some of them have been identified as edge states separating the other ones from the laminar state.

\end{itemize}

A standard way to characterize the subcritical transition to turbulence is to determine the average turbulent fraction of the flow as a function of the Reynolds number $R$, following either a localized perturbation or a quench in $R$.

For $R>R_g$, the turbulent fraction of the flow fluctuates around a given value, which remains non-zero on experimental timescales.
For $R<R_u$, the turbulent fraction rapidly converges to zero, 
without displaying any long transient regime.
Between these two values, for $R_u<R<R_g$, the turbulent fraction
first decays to a finite value after the perturbation, before entering
a long quasi-stationary regime in which it fluctuates around a well-defined non-zero value, until eventually a large fluctuation makes it decay to zero.
As mentioned above, the lifetime of the disordered flow is exponentially distributed, with a mean value $\tau$.
Whether $\tau$ diverges for some finite value of the Reynolds number is still a matter of debate.
A review of the numerous experimental and numerical studies addressing this question is presented in~\cite{Dauchot2012}. Typically three functional forms have been proposed to describe the experimental data (1) $\tau/\tau_0 = \exp(R/R_0)$, (2) $\tau/\tau_0 = \left(\frac{R_c}{R_c-R}\right)^{\alpha}, \alpha>0$, (3) $\ln(\tau/\tau_0) = \lambda \exp(R/R_0)$.
For the sake of "completeness", and because it is relevant in the context of the glass transition, where it is known as the Vogel-Fulcher-Tamman fit~\cite{Kivelson1996}, we have proposed a fourth one: (4) $\ln(\tau/\tau_0) = \lambda \left(\frac{R_c}{R_c-R}\right)^{\alpha}, \alpha>0$.  At first sight, one could think that such different functional dependencies could be easily discriminated from experimental data. However, as shown in~\cite{Dauchot2012}, even experimental data with up to five decades of timescales do not offer a clear distinction between the various functional forms, apart for the simple exponential one, which can be discarded.

In a recent work, Avila {\it et al}~\cite{Avila2011} report the super-exponential increase of the lifetimes of a \emph{single} turbulent patch. They then contrast it with the decrease of the typical time needed for one turbulent patch to split into two. These two time scales thus intersect at a given Reynolds number.  On average, below this Reynolds number patches relax before splitting and the turbulent state eventually relaxes towards the laminar flow. Above this Reynolds number patch splitting is strong enough to maintain the disordered flow on asymptotically long times. The authors thus conclude to the existence of a critical Reynolds number, while the timescale for the local relaxation has not diverged. In this view, it is the spatial coupling of transiently turbulent regions which gives rise to sustained turbulence.

\subsection{Spatiotemporal intermittency scenario}

The phenomenology reported above, and the importance of the spatial coupling had been emphasized, well before the aforementioned work, in the context of the study of spatiotemporal intermittency~\cite{Kaneko1985a,Chate1994,Chate1988}.  In a class of model systems called coupled map lattices, individual maps on a lattice, each of them exhibiting only transient chaos, still exhibit a phase transition towards a state with a non zero fraction of chaotic sites when the coupling among the maps is strong enough. Analogies to fluid flows had been already pointed out~\cite{Colovas1997,Bottin1998,Bottin1998b}. Also, the first author of this paper actually started his PhD thesis with Paul Manneville, studying the plane Couette flow as a prototype to investigate spatiotemporal intermittency, following a suggestion by Yves Pomeau~\cite{Pomeau1986}, in order to see whether it enters the universality class of directed percolation~\cite{Grassberger2006}. From that point of view the scenario of a transition to turbulence via spatiotemporal intermittency was considered as granted and there was no debate about the existence of an infinite lifetime turbulent state. Conversely, it is fair to say that whether this scenario applies to real flow had never been seriously questioned. At least the recent debate about the finite lifetimes of turbulence in shear flows has motivated the aforementioned work by Avila {\it et al}~\cite{Avila2011} and the one by Manneville {\it et al}~\cite{Manneville2009,Philip2011}, where system size effects have been investigated in details in the case of plane Couette flow.

All is well then? Not really. While it is reasonable, in the light of those recent works, to believe that the subcritical transition to turbulence enters the general class of transition via spatiotemporal intermittency, it remains that (i) the transition via spatiotemporal intermittency has itself no very solid theoretical grounds, (ii) the mapping of real flows on coupled map lattices remains a difficult task, which requires a lot of simplifications~\cite{Barkley2011}. From a theoretical point of view, one would like at least to answer a few questions such as : 
\begin{itemize}
\item What is the nature of the transition? Are there universal scaling properties to be expected, such as in other out-of-equilibrium phase transitions~\cite{Hinrichsen2000}? 
\item Is there a timescale that could be measured and which would diverge at the transition? If yes, what correlation length (if any) is it associated to?
\item What is the nature of the spatial coupling invoked in the spatiotemporal intermittency scenario? Are they local, diffusive-like, as in most coupled map lattice models, or global, as the pressure field is?
\item The low Reynolds number flows are reminiscent of the unstable finite amplitude solutions captured in experiments~\cite{Dauchot1995a,Bottin1998a} or in numerical simulations~\cite{Nagata1990,Clever1997,Cherhabili1996}. Together with their stable and unstable manifolds they form the skeleton of the disordered flow. At the transition the chaotic repellor associated with these solutions turns into a chaotic attractor. How is the transition related to the density and to the stability properties of these states in phase space? What is specific to the so-called edge-states~\cite{Skufca2006,DeLozar2012,Duguet2009,Schneider2010b,Schneider2007}, which separate the chaotic region in phase space from the laminar state?
\end{itemize}

\subsection{Complex phase-space picture and glass analogy}

As already mentioned above, the physics of glasses and of supercooled liquids shares interesting similarities with part of the phenomenology of the subcritical transition to turbulence; some of the questions raised are also quite similar in both fields.
For example, the fast increase of the relaxation time of a supercooled liquid, by many orders of magnitude, close to the glass transition temperature is often understood in the glass physics community as resulting from the wandering in a complex energy landscape of the phase-space point representing the system~\cite{bouchaud1997out}, this complex landscape being mostly composed of unstable fixed points~\cite{PhysRevLett.85.5356,PhysRevLett.85.5360,PhysRevLett.88.055502}.

Turning to real-space dynamics, another similarity between the glass transition and the transition to turbulence can be pointed out, namely the heterogeneity of the dynamics. In glasses, slowly evolving regions coexist with rapidly relaxing ones according to a complex spatiotemporal dynamics~\cite{vanSaarloos:2011wv}. Similarly, subcritical transitional flows also follow a complicated spatiotemporal dynamics, where fluctuating turbulent domains coexist with quiet laminar regions.
Interestingly, some of the one-dimensional models of glassy dynamics, called kinetically constrained models~\cite{Chandler:2010ws,ritort2003gdk}, form spatiotemporal patterns that are reminiscent of those observed in one-dimensional models exhibiting spatiotemporal intermittency~\cite{Chate1994,Barkley2011}. This analogy can even be made quantitative in some cases, in the sense that for some kinetically constrained models, the zero-temperature critical point belongs to the directed percolation class \cite{Chandler:2010ws}.

Beyond these analogies, let us also outline some important differences between the subcritical transition to turbulence and the glass transition.
A first clear distinction is that supercooled liquids are at equilibrium, whereas flows are maintained in a nonequilibrium state by an external driving.
A second important difference is that in subcritical flows, the laminar state is an absorbing state, from which turbulence cannot spontaneously emerge,
and the turbulence lifetime is defined as the time to relax to this absorbing state.
In contrast, the relaxation time in supercooled liquids is determined from the relaxation of stresses or density correlations and involves no absorbing state
--although one could think of the crystal as such a state.
Given these important differences, it is clear that one should not hope for a precise mapping between glasses and subcritical flows.
Rather, the goal of the present work is to consider the field of glasses and supercooled liquids, which is already well-developed at the methodological and theoretical levels, as a possible source of inspiration in the study of the subcritical transition to turbulence, given the similarities between some of the conceptual difficulties encountered in both fields.

\section{Glass transition: key theories and concepts}

Trying to get inspiration from a field at least as complex as the one, one is searching inspiration for, may seem a bit hazardous. At the same time, it might be the minimal condition to avoid too simplistic descriptions, inherited from low-dimensional dynamical systems. As a matter of fact, one sees from the above considerations, that dealing with spatially extended, fluctuating and disordered systems, most certainly requires some form of statistical analysis. As we shall see now, by introducing the reader to (some of) the theories of the glass transition, this is precisely the kind of situation such theories try to cope with. As already stated the reader should not expect that we will cover such a broad topic here. There are a number of review papers which do it in a very detailed way (see~\cite{Gotze1991a,Debenedetti1997,Donth2001,Debenedetti2001,Cavagna2009,Biroli2010,Berthier2011b}, to quote only a few) among which we particularly recommend~\cite{Debenedetti2001,Cavagna2009} to start with, and~\cite{Biroli2010,Berthier2011b} for the most recent developments. The goal of the next section is rather to give some entrance points, a taste of the concepts developed in that field, and to call attention to the way progress has been made.

\subsection{What is a glass?}

When a liquid is cooled down fast enough, in such a way that crystallization is avoided, it enters the so-called supercooled regime, which is a metastable state --in the sense that the crystal (when it exists) is the only thermodynamically stable state--, although it is equilibrated in the sense that time translational invariance holds. 
Decreasing further temperature, the viscosity, or equivalently the relaxation time of the liquid increases by more than ten orders of magnitude, while temperature is only decreased by a factor of two or three. At some point, the relaxation time exceeds the experimental timescale and the liquid falls out of equilibrium, namely time translational invariance is broken. The system ages: it has become a glass.

At the glass transition, one experimentally observes a sharp drop of the constant pressure specific heat, to a value very close to that of the crystalline phase. The above observation can be interpreted as the fact that in a glass, particles vibrate around their (disordered) equilibrium positions, with almost no structural rearrangement exactly as they would do around their ordered (equilibrium) positions in a crystal.  In that sense, the glass is an "amorphous crystal". The specific heat counts the degrees of freedom explored by the system : its drop indicates that some translational degrees of freedom have been lost and ergodicity is broken. 

Leaving the glass, we now come back to the equilibrated supercooled liquids. For the so-called "strong" glass formers, the relaxation time increases exponentially with the inverse temperature $\beta=1/T$, following a simple Arrhenius behavior and the supercooled liquid can be considered simply as a highly viscous liquid. But many glass formers --the so-called "fragile" ones-- experience a much faster growth of the relaxation time, suggesting that something more dramatic happens when approaching the glass transition.  
A more direct evidence of this qualitative change of the dynamical properties of the liquid is provided by the behavior of dynamic correlation functions such as the shear relaxation function --the integral of which is precisely the viscosity-- or the temporal correlation of density fluctuations.
A remarkable feature is that the qualitative shape of such dynamic correlation functions changes significantly approaching the glass transition. In the high temperature liquid the relaxation is purely exponential with a single relaxation time. When lowering the temperature one observes the formation of a plateau, when plotting the correlation function versus the logarithm of time.  A similar time separation is observed when monitoring the mean square displacement of the particles composing the liquid. At high temperature, it crosses over from the short time ballistic regime, to the long time diffusive regime. On the approach to the glass transition a plateau also develops, during which the mean square displacement remains constant, before eventually entering the diffusive regime. This phenomena has suggested the image of the particles being trapped or caged by their neighbors.
However, at the same time the structure of the supercooled liquid is strikingly similar to that of the high temperature liquid, at least as one can judge from standard static correlation functions, such as the pair correlation function. Then why are the particles trapped at low temperature and not at larger ones? Where does the increase in timescale, and the separation of timescales comes from, if it is not related to the structure? Or is it related to it, but in a very complicated way? The fact that the glass is as amorphous as the liquid, at least at first sight, makes it very difficult to answer these questions. Most theories of glasses try, with different perspectives, to address the above issues, and in particular to identify the proper length scale(s) associated with the qualitative change in the dynamical correlation functions.

\subsection{A brief and  partial history}

The "main-stream" theories of the glass transition can be organized chronologically in the following way. In 1948, Kauzmann~\cite{Kauzmann1948} has proposed a thermodynamic interpretation to the glass transition. It is followed ten years later by the Adam-Gibbs-DiMarzio phenomenological model~\cite{Gibbs1958,Adam1965}, which, on the basis of real space considerations, proposes a relation between thermodynamical and dynamical quantities. Later, in 1969, Goldstein~\cite{Goldstein1969} introduced the concept of energy landscape, that is a phase space view of the transition. One then has to wait until 1984 to see the first attempt to derive a theory from first principles. This is what the Mode Coupling Theory~\cite{Gotze1992} aims at, but at the price of highly technical calculations and pretty much uncontrolled approximations. Mode Coupling is in essence computing dynamical quantities from the static properties of the system. In parallel, in the early 80s, spin-glass theories~\cite{M.Mezard1987} were developed within a rather different community. Spin-glasses have a priori little to do with supercooled liquids. They notably have quenched disorder, while liquids do not. Still, the question of possible analogies was pending. In 1987, soon after a class of mean-field spin-glass models exhibiting a discontinuous glass transition was introduced --notably the Random Energy Model (REM), the p-spin model and the Potts spin-glass model--, Kirkpatrick and Wolynes noted the correspondence between the dynamic spin-glass theory and the Mode Coupling theory for liquids. Finally in 1989, the same authors proposed the so-called mosaic theory, (also known as the Random First Order Theory --RFOT), which aimed at reintroducing the real space in the mean-field picture of the p-spin, in the same way one introduces nucleation to describe first order transitions beyond mean-field. Since then, important developments have occurred both in the direction of a more rigorous formulation of the theory and in the search of the relevant length scales, both static~\cite{biroli2013comparison} and dynamic~\cite{vanSaarloos:2011wv}, which could be experimentally and numerically measured, in order to put some constraints on the theory. These last developments and a critical assessment of RFOT can be found in~\cite{Biroli2010}.

\subsection{A hand-waving sketch of the glass transition}

Let us start by the low temperature regime, close to the glass transition, where the supercooled liquid falls off-equilibrium. The commonly accepted picture is the one proposed by Goldstein~\cite{Goldstein1969}, of an energy landscape with many local minima, corresponding to different amorphous structures. The dynamics is dominated by activation processes, the system jumping over the barriers to explore the phase space in an ergodic manner. The corresponding image in real space is that the system rearranges locally the configuration of a small number $n$ of particles. Within this picture it is easy to figure out two relaxation times, a short one corresponding to vibrations around the frozen amorphous structures and a long one corresponding to the exploration of the different minima of the energy landscape. The large relaxation time $\tau$ then obeys the Ahrrenius law $\tau \sim \exp (\Delta/k_B T)$, where the height of the barrier $\Delta$ is, say, proportional to the number of particles $n$ concerned by one local rearrangement. An estimation of $n$ can be obtained, following the Adam-Gibbs-DiMarzio argument~\cite{Gibbs1958,Adam1965}, which goes as follows. It is assumed that the $n$ particles have a constant number $\omega$ of local configurations, independent of $n$. Then the total number of configurations of the whole system of $N$ particles is $\Omega \sim \omega^{N/n}$, where $N/n$ is the number of independent regions containing $n$ particles.
The so-called configurational entropy reads $s_{\rm conf} = N^{-1} \log\Omega \sim n^{-1} \log\omega$. Hence the height of the barrier, assumed to be proportional to $n$, is inversely proportional to the configurational entropy.
In the energy landscape picture, this configurational entropy counts the number of minima. If at some finite temperature, referred to as the Kauzmann temperature $T_K$, the number of minima becomes small in the sense that the configurational entropy $s_{\rm conf}$ vanishes in the limit of large $N$
and a phase transition takes place at $T_K$, characterized by $s_{\rm conf}(T_K)=0$.
At the transition, the height of the barriers, and thereby the relaxation time, diverge. Under some assumptions, an estimation of the configurational entropy can be obtained from experimental data on the temperature-dependence of
the heat capacity, leading to $s_{\rm conf}(T)\sim (T-T_K)/T_K$.
Inserting this dependence in the Arrhenius law for the relaxation time, one obtains the so-called Vogel-Fulcher-Tammann law, one of the fit proposed to describe the relaxation time at low temperature (see also the discussion about fitting the turbulent lifetimes in~\cite{Dauchot2012}).

The above scenario however breaks down when thermal energy becomes of the order of the heights of the energy barriers. It should also correspond to the situation where local rearrangements become less and less independent, an implicit hypothesis in the above estimate of the relaxation time. A naive thought would be that at such temperature, say $T_x$, the system simply crosses over to a standard liquid. This is however not the case: at the temperature where the above scenario breaks down, the liquid is still deeply supercooled, its viscosity being still $10^4$ to $10^5$ larger than that of the standard liquid, and most importantly the dynamical correlation functions still exhibit a significant plateau with two relaxation steps. 

The Mode Coupling theory precisely describes this high temperature regime. We shall certainly not enter the details of it, and it will be sufficient here to say that it predicts quantitatively well the shape of the dynamical correlations, but also that it predicts a divergence of the relaxation time in the form $\tau\sim (T-T_{MCT})^{-\gamma}$, with $T_{MCT}$ significantly larger than the glass transition temperature. This divergence is clearly not observed experimentally, but it is remarkable that $T_{MCT}$ coincides pretty well with the crossover temperature $T_x$.

The nature of this crossover was unveiled when the connection was made with the spin-glass models. Once it was established that the dynamical equation of the p-spin and the Mode Coupling ones were formally identical, one could take advantage of mean-field spin-glass models to compute many properties of the phase space, in particular the depth and the stiffness of the minima in the energy landscape. It was then established that the crossover temperature corresponds in phase space to a topological transition where minima turn into saddles. Since the model is mean-field, the energy barriers are infinite in the thermodynamics limit and the transition corresponds to a strict ergodicity breaking.

The final step is to come back to real liquids in finite dimension. One expects that here also the saddles and the minima control the dynamics. In this case, identifying minima and saddles of the potential energy landscape is a difficult but possible task, leading to interesting insights~\cite{PhysRevLett.85.5356,PhysRevLett.85.5360,PhysRevLett.88.055502}.
Finally one would like to identify in real space these "local" rearrangements. Are they purely dynamical events, or do they have a structural counterpart? We shall refer to the most recent literature \cite{Biroli2010} for those who are interested in this issue, and now comment on what we have learnt from this short introduction to the theories of the glass transition.

We see three points of interest. First, the reader may have noticed the similarity of the issue, between the glass transition and the subcritical transition to turbulence, regarding the existence of a divergence of the relaxation time
--even if relaxation means conceptually different things in both situations.
Glass physicists have tried for many years to fit experimental data with the same kind of laws as those we have introduced above
(apart from the simple exponential which here also could be discarded).
Second, in the glass transition also, it is the structure of phase space which governs the relaxation time. The spectrum of the saddles and the properties of the minima are key elements of the theories. Finally, and from our perspective it is the most important remark, the most significant progresses did not come from brute force calculations, starting from the Hamiltonian of the liquid. On the contrary, our understanding of the glass transition has greatly benefited from the study of oversimplified models, like the p-spin model or the Random Energy Model~\cite{PhysRevLett.45.79,PhysRevB.24.2613}, which describe the statistical behavior of a system evolving in a random energy landscape. 

In the next section, we briefly recall how we have tried in Ref.~\cite{Dauchot2012} to transpose the Random Energy Model to the context of the subcritical transition to turbulence, with the hope to gain insight into some possible underlying mechanisms of statistical nature which could play a role in this transition.

\section{A model inspired by the Random Energy Model}

Inspired by the analogy with glasses, the first step is to characterize the statistical properties of the energy landscape, in terms of the number of unstable solutions as a function of the turbulent energy density, for any Reynolds number.
Denoting $V$ the volume of the flow, the number of unstable solutions
at a given energy density $\ve=E/V$ is assumed to scale exponentially with the volume, $\Omega_V(\ve,R)\sim e^{V s(\ve,R)}$, leading to the definition of an entropy density $s(\ve,R)$.

As no unstable states exist for $R<R_u$, the entropy $s(\ve,R)$ is equal to zero in this regime. In the opposite case $R > R_u$, unstable states are assumed to exist only in the energy interval $(\ve_{\min}, \ve_{\max})$, implying that
the entropy is assumed to be nonzero only over an interval $\ve_{\min}(R) < \ve < \ve_{\max}(R)$.
From a dynamical viewpoint, experimental and numerical observations indicate that the turbulent flow spends a significant fraction of time in the vicinity of unstable solutions. At a coarse-grained level, the evolution of the flow can thus be represented as transitions between unstable solutions.
Altogether, the dynamics can be modeled as a diffusive process in energy space, biased by the entropy $s(\ve,R)$ to take into account the influence of the number of unstable states (see \cite{Dauchot2012} for details). Finally the presence of the absorbing laminar state is taken into account by assuming that when the flow ends up in the laminar state, the evolution stops until an external perturbation is imposed. 

To study the behavior of the model, we proceed in two steps. We first assume that the number of paths from the unstable states to the laminar one is small enough, so that the system essentially equilibrates within the unstable states before reaching the laminar state. A first step is thus to determine the equilibrium distribution in the absence of laminar state. The resulting equilibrium state exhibits a transition similar to the glass transition the REM.
Defining the turbulent lifetime as the mean time to reach the laminar state after a sudden quench from a higher Reynolds number value, where turbulence is established, the second step consists in computing the mean first passage time at the absorbing boundary $\ve=\ve_{\min}$.

Technically, the model is described by a Langevin equation for the energy $\ve$, and by the associated Fokker-Planck equation describing the evolution of the probability distribution $P(\ve,R,t)$.
Assuming reflecting boundary conditions at $\ve_{\min}$ and $\ve_{\max}$ to ensure the existence of an equilibrium state, the stationary distribution reads
\be \label{eq-dist-REM}
P(\ve,R) = \frac{1}{Z}\, \exp[V(s(\ve,R)-\beta(R) \ve)]
\ee
where $\beta(R)$ is a phenomenological parameter, akin to an inverse temperature, which characterizes the balance between energy injection and dissipation.

\begin{figure}[t!] 
\center
\vspace{-0.0cm}
\includegraphics[width = 0.95\columnwidth, trim = 0mm 0mm 0mm 0mm, clip]{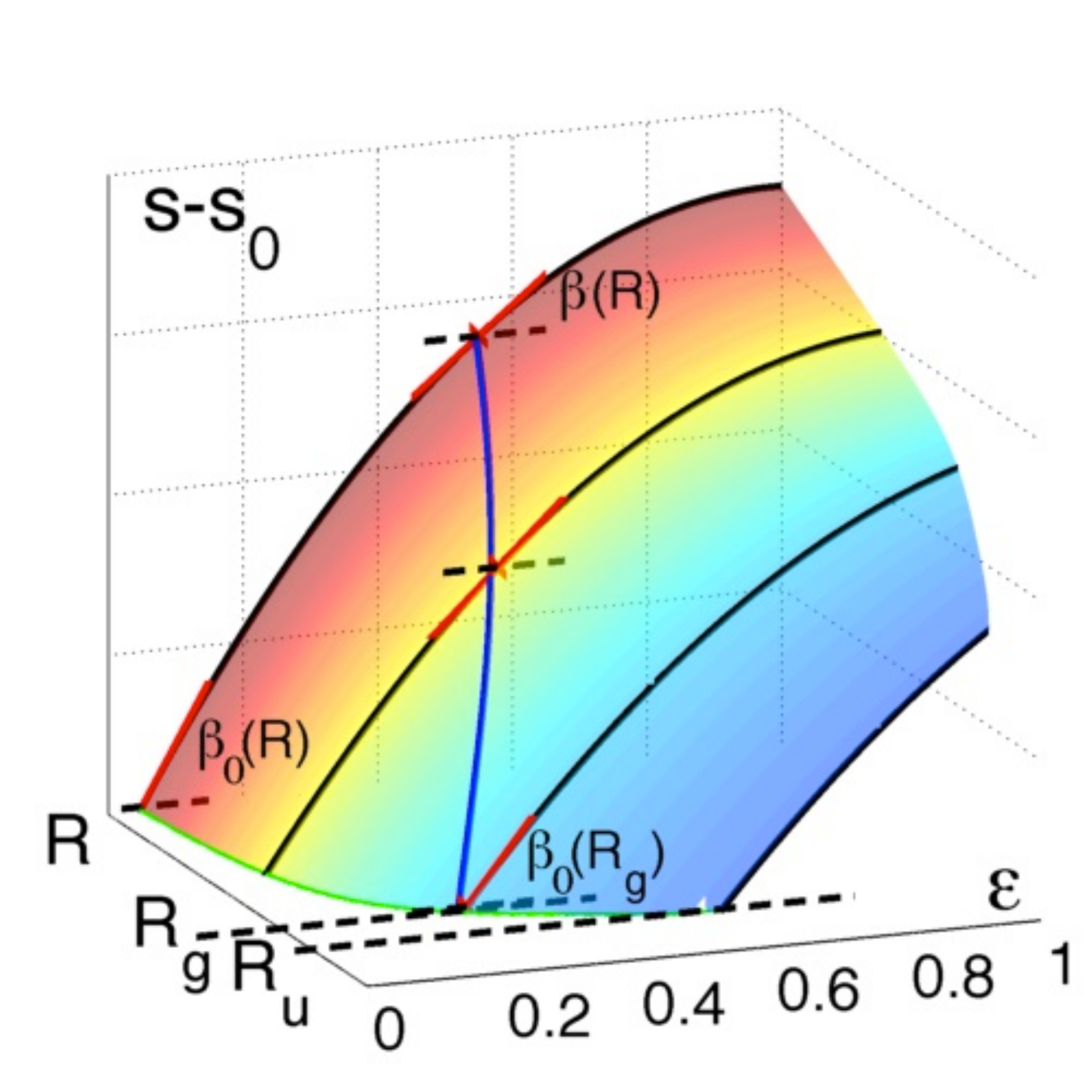}
\caption{(Color online) Schematic representation of the entropy $s(\ve,R)$, shifted by the value $s_0=s(\ve_{\min}(R),R)$. 
The color code on the surface, as well as the four black lines, indicate constant values of the Reynolds number.
From a graphical viewpoint, solving Eq.~(\ref{eq:maxarg}) requires to find a point on one of the constant Reynolds curves (for instance, one of the black lines) with a slope equal to $\beta(R)$.
A solution can only be found when $\beta(R) < \beta_0(R)$, where
$\beta_0(R)$ is the slope at the bottom of the surface (green line).
The physically observed solution when varying $R$ is given by the blue line,
which ends up at the value $R_g$ defined by $\beta(R_g) = \beta_0(R_g)$.}
\label{fig:entropy}
\vspace{-0.0cm}
\end{figure}

Fixing the Reynolds number, $P(\ve,R)$ is dominated at large $V$ by the energy $\bar\ve(R)$ maximizing the expression $s(\ve,R)-\beta(R) \ve$.
Two cases can then be distinguished: either $\bar\ve(R)$ falls between
$\ve_{\min}(R)$ and $\ve_{\max}(R)$, in which case $\bar\ve(R)$
is solution of 
\be \label{eq:maxarg}
s'(\bar\ve(R),R)-\beta(R) = 0,
\ee
or $\bar\ve(R)$ lies outside this interval.
In this last case, assuming the entropy $s(\ve,R)$ to be concave with respect to $\ve$ (see figure~\ref{fig:entropy}), its derivative 
$s(\ve,R)$ with respect to $\ve$ is maximum for $\ve=\ve_{\min}(R)$.
One of the key assumptions of the model, by analogy with the Random Energy Model \cite{PhysRevLett.45.79,PhysRevB.24.2613}, is that the derivative $s'(\ve,R)$ takes a finite value $\beta_0(R)$ in the limit $\ve \to \ve_{\min}(R)$ --see \cite{Dauchot2012} for a physical discussion of this assumptions.

If $\beta(R) < \beta_0(R)$, one sees from Eq.~(\ref{eq:maxarg}) that $\bar\ve(R) > \ve_{\min}(R)$. In the opposite case, when $\beta(R) > \beta_0(R)$,
no solution of Eq.~(\ref{eq:maxarg}) can be found in the interval
$\ve_{\min}(R) < \ve < \ve_{\max}(R)$ and $s(\ve,R)-\beta(R)\ve$ takes its maximum value for $\ve=\ve_{\min}(R)$, around which the probability distribution $P(\ve,R)$ concentrates.
Assuming that $\beta(R)$ decreases and that there exists a Reynolds number $R_g$ such that $\beta(R_g) = \beta_0(R_g)$, one has $\bar\ve(R) > \ve_{\min}(R)$ for $R > R_g$, while $\bar\ve(R) = \ve_{\min}(R)$ for $R < R_g$, meaning that in this last case the dynamics concentrates in the unstable states of lowest energy.

We thus see that under the equilibrium assumption (that is, neglecting the presence of the laminar state), a phase transition similar to the glass transition of the REM takes place.
Now taking into account the existence of the laminar state, one can compute the turbulence lifetime using first passage time techniques, modeling the presence of the laminar state by an absorbing boundary condition at $\ve_{\min}$. A natural question is then to know whether the equilibrium, REM-like transition has consequences on the turbulent lifetime.
To model a quench from a highly turbulent state, we choose as initial condition 
$\ve(t=0)=\ve_{\max}$.
In some linear approximation for the form of the entropy (see~\cite{Dauchot2012} for details), one obtains an explicit formula for the turbulent lifetime:
\be \label{eq-tau}
\tau = \frac{V}{D_0}\, (\Delta\ve)^2 \, f\Big( V(\beta_g-\beta)\, \Delta\ve \Big)
\ee
with the notations $\beta_g=\beta(R_g)$ and $\Delta \ve = \ve_{\max}-\ve_{\min}$;
the function $f(x)$ is defined as $f(x) = ( e^x - 1 -x)/x^2$.
When $\beta_g \ne \beta$, or equivalently $R \ne R_g$, one can use
for large $V$ the asymptotic behavior of $f(x)$ when $x \to \pm \infty$,
which is given by $f(x) \sim 1/|x|$ for $x \to -\infty$ and
$f(x) \sim e^x/x^2$ for $x \to +\infty$.
As a result, $\tau$ takes for $\beta_g< \beta$ the simple expression
\be \label{eq:tau:less}
\tau \sim \frac{\Delta \ve}{D_0 (\beta-\beta_g)} \, ,
\ee
which is seen to be independent of the volume $V$.
One thus finds a power-law divergence of $\tau$ as a function of the Reynolds number close to (but below) $R_g$,
\be \label{power:law:div}
\tau \sim \frac{\tau_0}{R_g-R} \, .
\ee
For a large but finite volume $V$, the divergence is smoothed very close to $R_g$, and the expression of $\tau$ crosses over to an exponential form, obtained for $R > R_g$,
\be \label{eq:tau:more}
\tau \sim \frac{e^{V(\beta_g-\beta) \Delta \ve}}{D_0 V (\beta_g-\beta)^2} \, .
\ee
At odds with Eq.~(\ref{eq:tau:less}), the volume $V$ does not disappear from the expression (\ref{eq:tau:more}) of $\tau$.
Only for an infinite volume $V$ does one recover a pure power-law divergence
for $R < R_g$.

\section{Conclusion}

The first goal of the present paper was to introduce hydrodynamicists to the physics of the glass transition, being motivated by the similarities we had noticed between the physics of the supercooled liquids and the one of the subcritical transition to turbulence.

As already stressed, those similarities rather sit, to some extent, at the conceptual level than at the phenomenological level. It seemed to us that some of the tools and concepts of the statistical physics of disordered systems, such as those developed in the context of the study of the glass transition, could be of interest to those interested in the subcritical transition to turbulence.

Being aware that the theories of the glass transition are a vast, and sometime opaque field of research, we have tried to bring our colleagues a very schematic presentation of both the history and the main concepts underlying the development of this field. The price to pay for such a partial presentation is obviously to remain at a superficial level of description. This is why we have tried to provide, within the bibliography, both original papers and reviews, which could help the interested reader to enter deeper into this fascinating topic.

Finally, we have summarized the main ingredients and findings of a model inspired by the Random Energy Model, which we had recently proposed. This simplistic model leads to an estimate of the turbulence lifetime as a function of the Reynolds number close to the transition, an estimate which qualitatively agrees amazingly well with the observed phenomenology. We hope that our results will motivate more involved studies in that same spirit.

\bibliographystyle{plain}
\bibliography{Turbulence,Glasses,Glasses2}

\end{document}